\documentclass[twocolumn,twoside,slac_two]{revtex4}
\usepackage{graphicx}
\usepackage{fancyhdr}
\begin{document}

\title{Connection between gamma-ray and radio activity of blazars from \emph{Fermi}-LAT and 15~GHz radio monitoring with the OVRO 40-Meter Telescope}

%

\author{W. Max-Moerbeck on behalf of the \emph{Fermi} Large Area Telescope Collaboration and the F-GAMMA Collaboration}
\affiliation{Department of Astronomy, California Institute of Technology. Pasadena, CA 91125, USA}
%
%

\begin{abstract}
A large sample of blazar from the Candidate Gamma Ray Blazar Survey (CGRaBS) has been observed with the Owens Valley Radio Observatory (OVRO) 40-Meter Telescope at 15~GHz. Using these quasi-simultaneous observations, we study the connection between the gamma-ray behavior of blazars as detected by \emph{Fermi}-LAT and the cm band as observed by the F-GAMMA project with the OVRO 40-Meter Telescope. Comparing the light curves for a large number of sources, it is possible to study in detail the relation between the gamma-ray and radio activity of \emph{Fermi}-LAT detected gamma-ray blazars. We present first results for correlations between \emph{Fermi}-LAT and our 15~GHz observations.
\end{abstract}

\maketitle

\thispagestyle{fancy}


\section{INTRODUCTION}
Blazars are a subclass of active galactic nuclei characterized by rapid variability, strong linear polarization at visible wavelengths, apparent superluminal motions, flat radio spectrum and double-peaked spectral energy density. They are further divided into two types: BL Lacs and Flat Spectrum Radio Quasars (FSRQs). BL Lacs have almost no emission lines, strong polarization and are preferentially found in elliptical galaxies. FSRQs have broad and narrow emission lines, strong polarization and are more luminous than BL Lacs. It is believed that all AGN are powered by an accreting super-massive black hole and that the different classes, blazars included, are due to differences in the observing angle and the presence or absence of jets. In this picture blazars would be objects with a jet oriented close to the line of sight.

Given the broad-band spectral energy distributions and the variability of these objects, a complete understanding of them requires multiwavelength and multi-epoch data. The Large Area Telescope (LAT) on \emph{Fermi} Gamma-ray Space Telescope provides gamma-ray light curves for the brightest objects. These can be combined with our radio data and other data sets in order to investigate the relation between blazar properties at different energy bands. 

A statistically large sample with high-cadence is required to answer questions about the relation between gamma-ray and radio emission. The radio data by itself can be used to characterize the variability properties of these large sample and study their variations among different classes. Questions that can be addressed are for example: Are all blazars equally variable on long time scales? Different populations or smooth variation over the properties? What are the characteristic time scales? Are there correlations between variability characteristics and physical properties of central black hole and/or jet.

To address some of these issues we have embarked on an extensive monitoring program in the radio band. The OVRO 40-Meter Telescope monitoring program is part of the \emph{Fermi}-GST Multi-wavelength Monitoring Alliance (F-GAMMA) project~\citep{fuhrmann_simultaneous_2007,angelakis_monitoringradio_2008}. The other component of the project consist of monthly multifrequency monitoring of 60 sources selected by historically interesting behavior. Observations are carried out with the Effelsberg 100~m and Pico Veleta 30~m telescopes at 12 frequencies between 2.7 and 270~GHz. Combination of these two programs allows detailed study of the spectral evolution for the small sample, and population studies for the large high-cadence single frequency sample.

Here we present first results for the study of correlated variability using \emph{Fermi}-LAT observations and radio light curves from our monitoring program.

\section{OBSERVATIONS}
Gamma-ray light curves for the brightest sources can be obtained from \emph{Fermi}-LAT observations. This has been done on weekly time scales for the source on the LBAS sample \citep{abdo_bright_2009} using the first 11 months of \emph{Fermi}-LAT observations. Details on the \emph{Fermi}-LAT instrument and the data reduction can be found in other articles in these conference proceedings or elsewhere.

Observations of all the sources with $\delta > -20^{\circ}$ from the Candidate Gamma Ray Blazar Survey (CRGaBS) \citep{healey_cgrabs:all-sky_2008} have been carried out using the OVRO 40-Meter Telescope at 15~GHz. The sources have been observed about twice a week since mid-2007 with a thermal noise floor of about 5~mJy. Observations are made using a dual-beam Dicke-switch system. The individual beams are 2.5' FWHM; they are separated by 13'. The cryogenic HEMT receiver has a noise temperature of $\sim$30 K and covers the band from 13.5 to 16.5~GHz. Results based on the radio data only are presented by Richards et al. [2009] on this conference proceedings.

\section{FIRST RESULTS}
One of the basic questions we want to explore is the presence of simultaneous variability in radio and gamma-ray bands. The first step is to determine the presence of variability for the sources in both bands. Using a $\chi^2$ test to evaluate the probability of a constant source producing the observations, it is found that $\sim75\%$ of the sources are variable at the 99.9\% confidence level in both bands and only $\sim6\%$ are not variable in either band. The rest of the sources ($\sim19\%$) are variable in one of the two bands. The basic observation is that objects that vary at gamma-ray wavelengths are almost always also variable at radio wavelengths over period of time included in our observations. The few exceptions are probably due to sampling, duty cycle, low level of variability effects among other possible causes.

To address the question of the timing between the activity at the two bands we have used cross-correlations. These cross-correlation are investigated using the method described in \citet{edelson_1988}. This method allows the use of unevenly sampled light curves such as the ones used in this study, but doesn't provide a quantification of the significance of the observed correlations. This issue will be addressed in a publication in preparation.

\begin{figure*}[!ht]
\centering
\includegraphics[width=170mm]{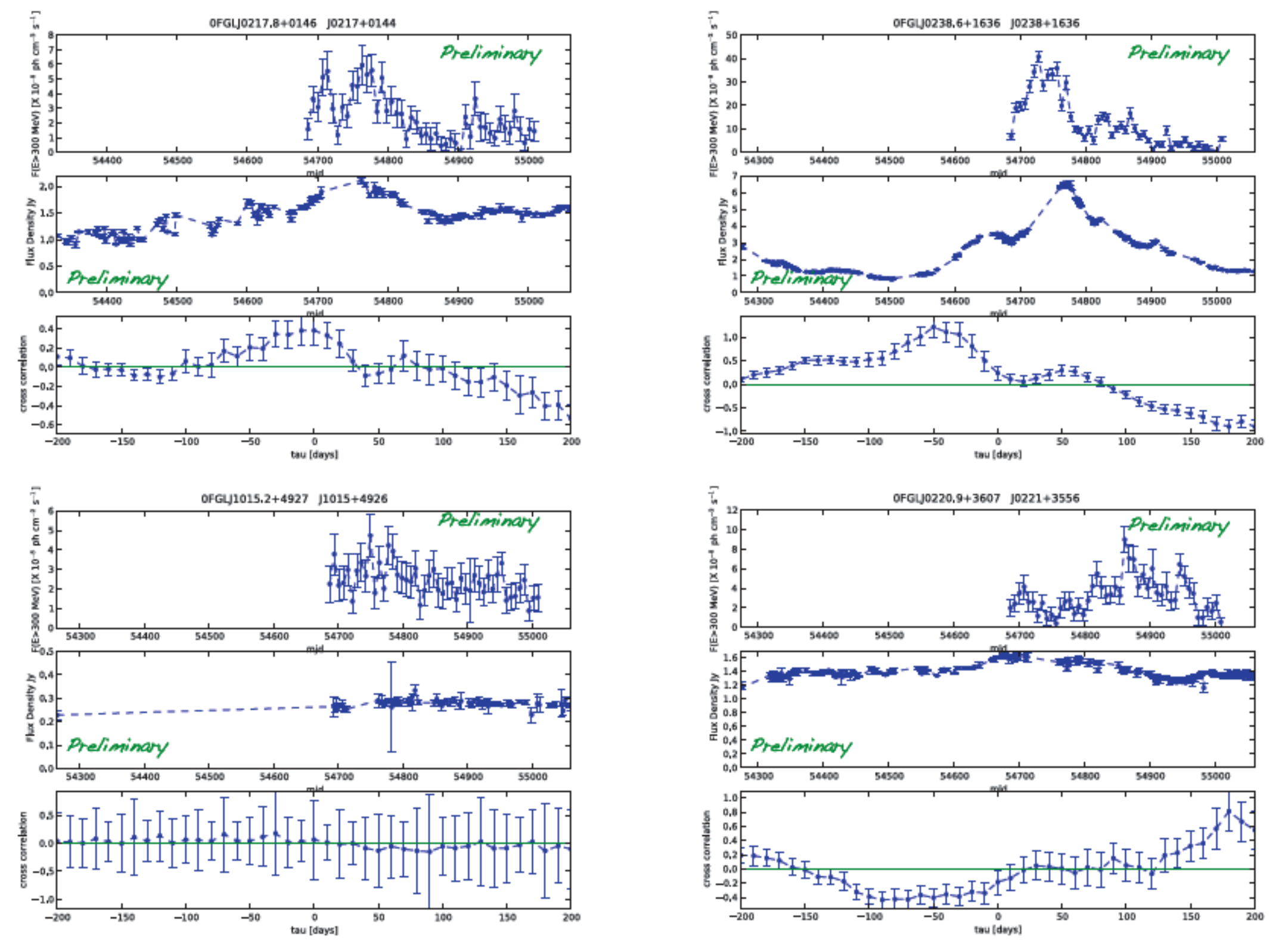}
\caption{Examples of gamma-ray, radio light curves and cross-correlations for four sources included in this study. The data reduction and cross-correlations are preliminary, and are included to show the quality of the data and the variety of observed behaviours.} \label{fig:example_data}
\end{figure*}

We include some of our cross-correlation results here to give an example of the quality of the data on the two bands (Fig. \ref{fig:example_data}). In view of the fact that the flares in these plots often last for a significant fraction of the duration of the observations, it is not surprising that high cross-correlations are seen. This does not, however, indicate that these events are related at the two frequencies. Given the short duration of the light curves, it is not yet possible to determine if the single observed gamma-ray flare is related to the observed radio flare, or to one occurring before/after and not included in the observations. Careful statistical analysis and longer duration observations are needed before any statistical significance can be attached to these correlations.

\section{CONCLUSIONS}
 It is clear from the observations that very interesting behavior occurs at both frequencies, but it is premature at this stage to try to associate specific features seen at gamma-ray frequencies with features seen at radio frequencies, or vice versa, or to claim that a specific gamma-ray feature leads, or lags, a specific radio feature. Longer term monitoring and careful statistical analyses are required before we can make any such definitive statements. A publication addressing these issues is in preparation.

\bigskip 
\begin{acknowledgments}
The \textit{Fermi} LAT Collaboration acknowledges generous ongoing support from a number of agencies and institutes that have supported both the development and the operation of the LAT as well as scientific data analysis. These include the National Aeronautics and Space Administration and the Department of Energy in the United States, the Commissariat \`a l'Energie Atomique and the Centre National de la Recherche Scientifique / Institut National de Physique Nucl\'eaire et de Physique des Particules in France, the Agenzia Spaziale Italiana and the Istituto Nazionale di Fisica Nucleare in Italy, the Ministry of Education, Culture, Sports, Science and Technology (MEXT), High Energy Accelerator Research Organization (KEK) and Japan Aerospace Exploration Agency (JAXA) in Japan, and the K.~A.~Wallenberg Foundation, the Swedish Research Council and the Swedish National Space Board in Sweden.

Additional support for science analysis during the operations phase is gratefully acknowledged from the Istituto Nazionale di Astrofisica in Italy and the Centre National d'\'Etudes Spatiales in France.

The Owens Valley Radio Observatory 40-Meter Telescope monitoring program is supported in part by NASA (grant NNX08AW31G) and the NSF (grant AST-0808050).

WM acknowledges support from the U.S. Department of State and the Comisi\'on Nacional de Investigaci\'on Cient\'ifica y Tecnologica (CONICYT) in Chile, for a Fulbright-CONICYT scholarship.
\end{acknowledgments}

\bigskip 
\bibliography{wmaxFermiSympBib}

\end{document}